\documentclass[letter,printer]{aa}  

\usepackage{graphicx}
\usepackage{txfonts}

\begin{document}

   \title{Spiral arms and instability within the AFGL 4176 mm1 disc}

   \author{Katharine G. Johnston \inst{1}
          \and
          Melvin G. Hoare \inst{1}
          \and 
          Henrik Beuther \inst{2}
          \and
          Rolf Kuiper \inst{3}
          \and
          Nathaniel Dylan Kee \inst{4}
          \and
          Hendrik Linz \inst{2}
          \and \\
          Paul Boley \inst{5,6}
          \and
          Luke T. Maud \inst{7}
          \and
          Aida Ahmadi \inst{2}
          \and
          Thomas P. Robitaille \inst{8}
          }

   \institute{School of Physics \& Astronomy, E.C. Stoner Building, The University of Leeds, Leeds, LS2 9JT, UK\\
              \email{k.g.johnston@leeds.ac.uk}
              \and
             Max Planck Institute for Astronomy, K\"onigstuhl 17, D-69117 Heidelberg, Germany
             \and
             Institute of Astronomy and Astrophysics, Eberhard Karls University T\"ubingen, Auf der Morgenstelle 10, D-72076 T\"ubingen, Germany
             \and
             Institute of Astronomy, KU Leuven, Celestijnenlaan 200D, B-3001 Leuven, Belgium
             \and
             Moscow Institute of Physics and Technology, 9 Institutskiy per., Dolgoprudny 141701, Russia
             \and
             Ural Federal University, 19 Mira st., Ekaterinburg 620075, Russia
             \and
             ESO Headquarters, Karl-Schwarzschild-Str. 2, 85748 Garching bei M\"unchen, Germany
             \and
             Aperio Software, Headingley Enterprise and Arts Centre, Bennett Road, Headingley, Leeds, LS6 3HN, UK
             }

  \date{Received 20 November 2019 / Accepted 21 January 2020}
 
  \abstract{We present high-resolution (30\,mas or 130\,au at 4.2\,kpc) Atacama Large Millimeter/submillimeter Array observations at 1.2\,mm of the disc around the forming O-type star \object{AFGL 4176} mm1. The disc (AFGL\,4176 mm1-main) has a radius of $\sim$1000\,au and contains significant structure, most notably a spiral arm on its redshifted side. We fitted the observed spiral with logarithmic and Archimedean spiral models. We find that both models can describe its structure, but the Archimedean spiral with a varying pitch angle fits its morphology marginally better. As well as signatures of rotation across the disc, we observe gas arcs in CH$_3$CN that connect to other millimetre continuum sources in the field, supporting the picture of interactions within a small cluster around AFGL\,4176 mm1-main. Using local thermodynamic equilibrium modelling of the CH$_3$CN K-ladder, we determine the temperature and velocity field across the disc, and thus produce a map of the Toomre stability parameter. Our results indicate that the outer disc is gravitationally unstable and has already fragmented or is likely to fragment in the future, possibly producing further companions. These observations provide evidence that disc fragmentation is one possible pathway towards explaining the high fraction of multiple systems around high-mass stars.}


   \maketitle

\section{Introduction \label{intro}}

Recently, the Atacama Large Millimeter/submillimeter Array (ALMA) has detected spiral structures in the midplanes of both protostellar \citep[e.g.][]{tobin16a,lee19a} and protoplanetary discs \citep[e.g.][]{perez16a,huang18a,kurtovic18a} around low-mass  ($<2\,M_{\odot}$) stars. In this Letter, we present observations which show spiral structure within the disc of the forming massive star AFGL\,4176 mm1, which lies at a distance of 4.2\,kpc \citep{green11a}. This result accompanies recent observations of spiral structure in the discs or inner envelopes around other massive stars \citep[e.g.][]{csengeri18a,maud19a}.

We previously presented ALMA observations of the AFGL\,4176 region, uncovering one of the best examples of a near-Keplerian disc around an O-type star \citep[][hereafter J15]{johnston15a}. Radiative transfer modelling by J15 of these data determined that the best-fitting disc mass and radius were 12\,M$_{\odot}$ and 2000\,au, respectively, and the inclination was close to face-on ($i=30^{\circ}$). 

Spiral structure in discs around young high-mass stars has been anticipated by a host of simulations \citep[e.g.][]{krumholz07e, kuiper11, harries17a, meyer18a, ahmadi19a}, which predict near-Keplerian discs with spiral arms, sizes of the order of hundreds to thousands of astronomical units, and masses of several $\sim$10\% of their host stars. In some cases, disc fragments are predicted \citep[e.g.][]{meyer18a, ahmadi19a, rosen19a}, which may have also been observed recently for the first time \citep{ilee18a}.

The spiral structure seen in these simulations stems mainly from their high disc masses, producing discs that are gravitationally unstable. The Toomre parameter $Q$ \citep{toomre64} can be used as a measure of the level of gravitational instability within a disc: 
\begin{equation}
Q = \frac{c_s \kappa}{\pi G \Sigma} 
\label{toomre_eqn}
\end{equation}
where $c_s$ is the sound speed, $\kappa$ is the epicyclic frequency, $G$ is the gravitational constant, and $\Sigma$ is the disc surface density. A range of simulations have shown that values of Q$<$1.5-1.7 are required for gravitational instability to occur \citep[][and references therein]{durisen07a,durisen08a}.

The importance of gravitational instabilities in high-mass star formation are far-reaching as they likely play a role in observed variability due to accretion bursts \citep[e.g.][]{caratti-o-garatti17a, hunter17a}. Additionally, they have the potential to explain the high binary and multiple fraction for high-mass stars \citep{duchene13a}.

\section{Observations \label{observations}}

We observed AFGL\,4176 with the 12\,m antenna array of ALMA Cycle 5 under program 2017.1.00700.S (PI: Johnston). The observations were taken in dual polarisation mode in Band 6 at a frequency of $\sim$250\,GHz (1.2\,mm), with one pointing centred on 13$^{h}$43$^{m}$01.689$^{s}$ $-$62$^{\circ}$08$'$51.25$''$ (ICRS). The observations were taken in three different configurations on 3 October 2017, 4 November 2017, and 25 November 2017, with baseline lengths ranging between 118.1\,m-15.0\,km, 113.0\,m-13.9\,km, and 92.1\,m-8.5\,km, and measurements of precipitable water vapour of 1.1, 0.5, and 0.6\,mm on each date, respectively. Therefore, the angular resolution for the continuum is $\sim$30\,mas or $\sim$130\,au (d=4.2\,kpc, robust=0.5 weighting) and the largest angular scale that the observations are sensitive to is $\sim$1.6$''$ (or 6700\,au, corresponding to a baseline of 92.1\,m). The number of antennas for each observation was 42, 43, and 44 (with 30, 41, and 43 having useful data, respectively). The majority of the 12 antennas removed from the first observation were flagged during the pipeline reduction due to poor phase stability because of high wind. The primary beam sizes range between 20.5 and 22.0$''$. The spectral setup is described in Table~\ref{spw_setup}. The bandpass and absolute flux calibrator was J1427-4206, the phase calibrator was J1337-6509, the check source was J1308-6707, and the pointing calibrator was J1424-6807.

The data reduction was carried out using Common Astronomy Software Applications \citep[CASA, ][]{mcmullin07a} version 5.1.1-5 r40000 via the pipeline version 40896 (Pipeline-CASA51-P2-B), and imaging was carried out in CASA version 5.3.0. In addition to the pipeline calibration, we performed phase-only self-calibration to the continuum and applied these solutions to the line data before imaging. The continuum was imaged using multiscale clean and a Briggs weighting parameter of 0.5. The central frequency of the combined continuum image made from the line-free channels is 248.057\,GHz (1.21\,mm). The line data presented in this Letter (K-ladder transitions of CH$_3$CN) were imaged using multiscale clean, a robust parameter of 1.0, and a channel width of 1\,km\,s$^{-1}$ to improve sensitivity.

\begin{table}
\caption{ALMA spectral setup.\label{spw_setup}}
\centering
\small
\begin{tabular}{llllll}
\hline \hline
Central Frequency & Bandwidth & Number of & \multicolumn{2}{c}{Spectral Resolution} \\
(GHz) & (GHz) & Channels & (kHz) & (km\,s$^{-1}$) \\
\hline
239.298 & 0.469 & 1920 & 282 & 0.353 \\
241.478 & 1.875 & 1920 & 1129 & 1.40 \\
253.106 & 1.875 & 1920 &1129 & 1.34 \\
255.411 & 1.875 & 1920 & 1129 & 1.33 \\
\hline
\end{tabular}
\tablefoot{The spectral resolution is given after Hanning smoothing.}
\end{table}

\section{Results and discussion \label{results}}

\subsection{1.2\,mm continuum \label{contin_sec}}

\begin{figure}
\centering
\includegraphics[width=9cm]{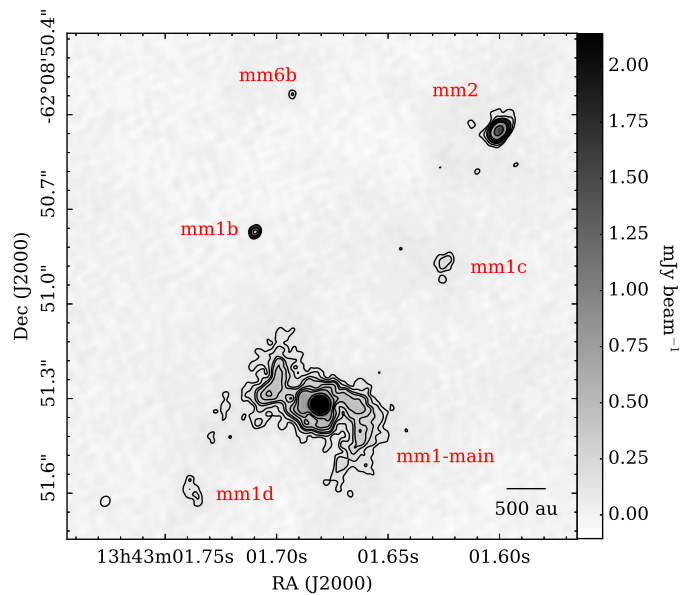}
  \caption{1.21\,mm continuum emission of the AFGL\,4176 region in greyscale and contours ($\sigma$ = 23\,$\mu$Jy\,beam$^{-1}$ $
 \times$ -5, 5, 7, 10, 12, 16, 20, 25, 50, 100). The beam is shown in the bottom left corner (0.034$'' \times$0.028$''$, PA=-33.5$^{\circ}$), and a scalebar is shown in the bottom right corner. The millimetre sources are labelled.}
     \label{contin}
\end{figure}

Figure~\ref{contin} shows the 1.21\,mm continuum emission from the AFGL\,4176 region. The new observations show that the source AFGL\,4176 mm1 presented in J15 now splits into four sources: mm1-main, mm1b, mm1c, and mm1d. The source mm2 is still single and shows an elongation, which extends in the direction away from mm1.

Most notably, the new observations show the disc source mm1-main has a significant amount of sub-structure. The west side of the disc contains a spiral arm which emanates from the north-west of the central region of the disc, whereas the east side resembles an arc-like structure which curves like a closing parenthesis away from the centre of the disc. \citet{meyer18a} present simulated ALMA 1.2\,mm continuum observations of a radiation-hydrodynamical simulation of a system with similar properties as AFGL\,4176 (their Fig. 16). Compared to our Fig.~\ref{contin}, there are several similarities. If the eastern arc in our observations is interpreted as another spiral arm, which has a fragment forming in the southern part of the arc, both discs have an asymmetric two-armed morphology. In addition, the end of the north-eastern spiral arm in the simulated observation resembles the morphology of the eastern arc-like structure, with a peak or possible fragment in the south-east end of the arc. Thus, the morphology of the dust in our observations is consistent with that seen in simulations of gravitationally unstable discs around forming massive stars.

The peak flux of mm1-main is 10.72$\pm$0.02\,mJy\,beam$^{-1}$ at a position of 13:43:01.681 $-$62:08:51.32 (ICRS), and the integrated flux above 1$\sigma$ is 42.67$\pm$0.03\,mJy. The integrated flux for mm1 is slightly smaller than that determined by J15 with a resolution of $\sim$0.3$''$ (50$\pm$4\,mJy), but it is still within 2$\sigma$. Using the same assumptions as J15, including $T=$190\,K, $\kappa_\nu=$0.24\,cm$^{2}$g$^{-1}$, and a total-mass-to-dust ratio of 154, we find that the disc mass is 6.6\,M$_{\odot}$, which should be compared to the mass of $\sim$8\,M$_{\odot}$ determined by J15. We note however that full radiative-transfer modelling in J15 determined a disc mass of 12\,M$_{\odot}$. 

There is a central bright compact source at the centre of the mm1 disc. We fitted this central source with a 2D Gaussian using imfit in CASA and subtracted it, thus finding a peak flux of 10.4\,mJy\,beam$^{-1}$, an integrated flux of 13.9\,mJy, as well as convolved and deconvolved sizes of 35.6$\times$ 35.5\,mas, PA=55$^{\circ}$ and 21.7$\times$11.6\,mas (corresponding to 91$\times$49\,au), PA=56$^{\circ}$, respectively. Within the positional accuracy of the Australia Telescope Compact Array (ATCA; $\sim$0.3$''$), there is a coincident compact continuum source at 1.23\,cm measured with ATCA with an integrated flux of 1.33\,mJy (Johnston et al., submitted). Therefore, the spectral index between the centimetre and millimetre compact continuum sources is $\sim$0.9. If the compact sources at both wavelengths are entirely due to ionised gas, then this would indicate there is compact, partially-optically-thick ionised gas associated with the centre of the disc, which may be a jet or a hypercompact HII region. In addition, removal of the compact source from the 1.21\,mm emission would reduce the total disc mass from 6.6 to 4.4\,M$_{\odot}$. However, we note that a fraction of the compact millimetre source is likely also due to the rising surface density expected at the centre of the disc \citep[as seen in the simulated observation of ][]{meyer18a}.

By using an ellipse with PA=60$^{\circ}$, we measured the semi-major axis of mm1-main including the spiral arms at 5$\sigma$ to be $\sim$0.25$''$, giving a disc radius of $\sim$1000\,au. We also used the ratio of the semi-major-to-minor axes of the disc to determine an estimate of the disc inclination. The smallest semi-minor axis that is consistent with the shape of the disc is 0.15$''$ (this is illustrated in Fig.~\ref{ellipse}), providing an upper limit to the inclination of 51$^{\circ}$. This is consistent with the inclination of 30$^{\circ}$ found for the best-fitting model of J15.

\begin{figure}
\centering
\includegraphics[width=9cm]{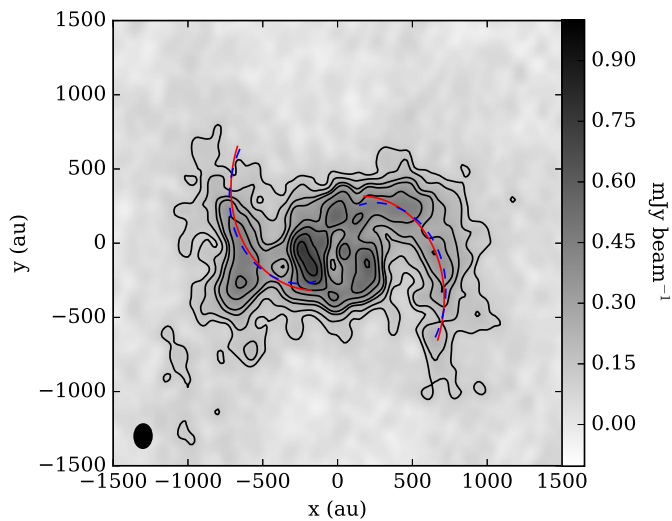}
  \caption{Deprojected 1.21\,mm continuum image of AFGL\,4176 mm1-main in greyscale and contours, with the central compact source subtracted (contours: $\sigma$ = 23\,$\mu$Jy\,beam$^{-1}$ $ \times$ -5, 5, 7, 10, 12, 16, 20, 25, 30). The rotated and deprojected beam is shown in the bottom left corner. The best fitting logarithmic and Archimedean spiral models (cf. Fig.~\ref{cutdisc}) are plotted as blue dashed and red solid lines on the western (right) spiral, respectively. The same models rotated 180$^{\circ}$ are also shown.}
     \label{zoomcont}
\end{figure}

\begin{figure}
\centering
\includegraphics[width=9.1cm]{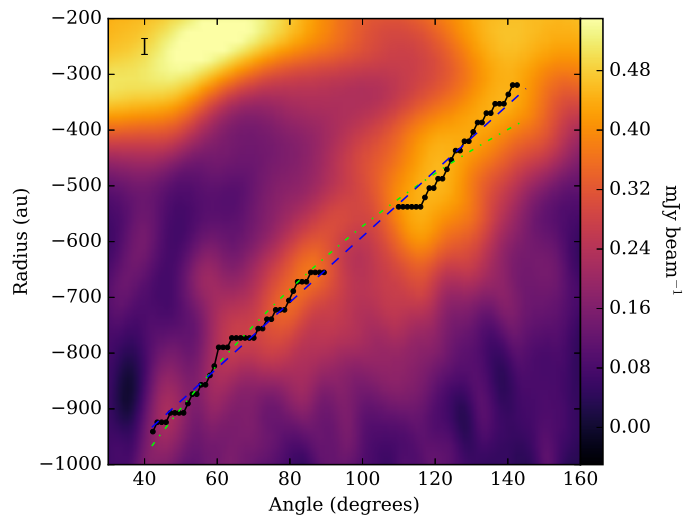}
  \caption{Flux density as a function of radius and position angle. The peak radius for each position angle is plotted as a black line and points for the western spiral arm, and the green dot-dashed and blue dashed lines show the best fits for logarithmic and Archimedean spiral models, respectively. A positional error bar is shown in the top left corner.}
     \label{cutdisc}
\end{figure}

In Fig.~\ref{zoomcont} we show the residual image of mm1-main. The residual image has the central compact source removed and is deprojected, assuming a position angle of 60$^{\circ}$ and an inclination angle of 30$^{\circ}$, which we assume in the remainder of this analysis. Figure~\ref{cutdisc} shows the same data as a function of radius from mm1-main and position angle around the disc. After masking Fig.~\ref{cutdisc} so that only the western (now right) spiral arm remains, we find the peak pixel and therefore radius for each position angle, which is plotted as a black line in Fig.~\ref{cutdisc}.

Similar to the analysis that was carried out for the Disk Substructures at High Angular Resolution Project \citep[DSHARP; e.g.][]{kurtovic18a, huang18a}, we fitted the peak positions of the western spiral arm with both logarithmic and Archimedean spirals using the \texttt{curve\_fit} function in \texttt{scipy} with the Levenberg-Marquardt algorithm.  We assume that the error in determining the peak position in radius is half the major axis of the beam divided by the lowest signal-to-noise ratio of the detected emission in the arm, that is, five \citep{taylor99a}. We also weighted each datapoint by the spacing in beams between the sampled position angles at each radius to account for correlation within the scale of a beam. The equations for the models, as a function of radius $r$ and position angle $\theta$ around the disc, and best fit parameters are given in Table~\ref{spiral_params}. With a lower $\chi^2$ of 8.9, the Archimedean spiral model more closely reproduces the data. 

\begin{table*}
\caption{Spiral models and associated fit parameters\label{spiral_params}}
\centering
\small
\begin{tabular}{lllllr}
\hline \hline
Spiral type & Equation & $r_0$ & $b$ & Pitch angle & $\chi^2$ \\
\hline
Logarithmic & $r = r_0 \exp{(b\theta)}$ & $-3184 \pm 164$\,au & $-0.5175 \pm 0.0185$ & $\arctan{(b)}= -27.4^{\circ}$ & 17.8 \\
Archimedean & $r = r_0 + b\theta$ & $-1712 \pm 32$\,au & $337.9 \pm 10.5$\,au & $\arctan{(b/r)} = -19.8 \rightarrow -46.6^{\circ}$ & 8.9 \\
\hline
\end{tabular}
\end{table*}

Symmetric, tightly-wound logarithmic spirals with a constant pitch angle are expected from simulations of gravitational instabilities (GI) in discs \citep[e.g.][]{cossins09a}, whereas asymmetric spirals with a varying pitch angle are expected in the case of companion-induced spiral arms \citep[e.g.][]{rafikov02a, forgan18d}. However, when these models are compared using simulated observations, spiral arms in discs produced by a companion can look similar to those predicted for GI \citep[e.g.][]{meru17a}. Nevertheless, the fact that there is not a clear symmetrical spiral pattern on the other side of the disc suggests that the spiral structure is generated to some degree by the existence of one or more nearby companions. In fact, the other continuum sources in the field seen in Fig.~\ref{contin} may act as these. These observed companions may have been produced by previous episodes of gravitational instability within the disc and later ejected, or formed from the surrounding cloud.

\subsection{CH$_3$CN line emission}

\begin{figure}
\centering
\includegraphics[width=9cm]{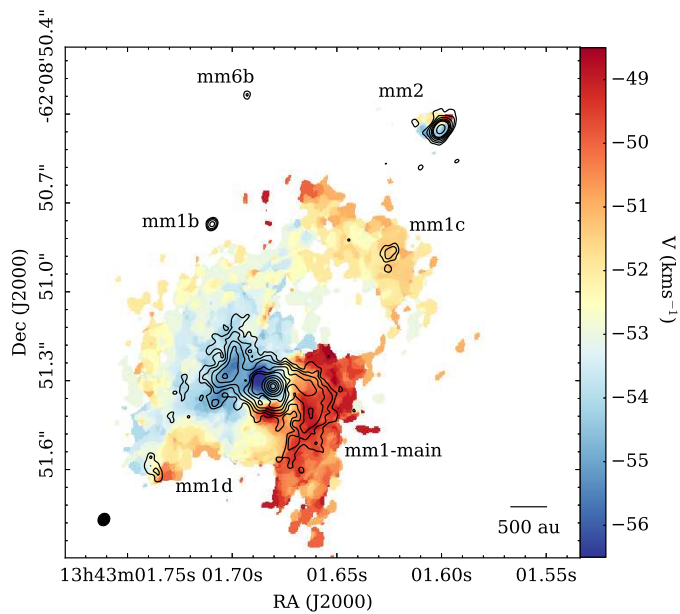}
  \caption{CH$_3$CN J=13-12 K=3 first moment map in colourscale and 1.21\,mm continuum in contours (same as in Fig.~\ref{contin}). The beam is shown in the bottom left corner (0.043$'' \times$0.038$''$, PA=-32.6$^{\circ}$), and a scalebar is shown in the bottom right corner. }
     \label{ch3cn_fig}
\end{figure}

\begin{figure*}
   \centering
            {\includegraphics[width=8.7cm]{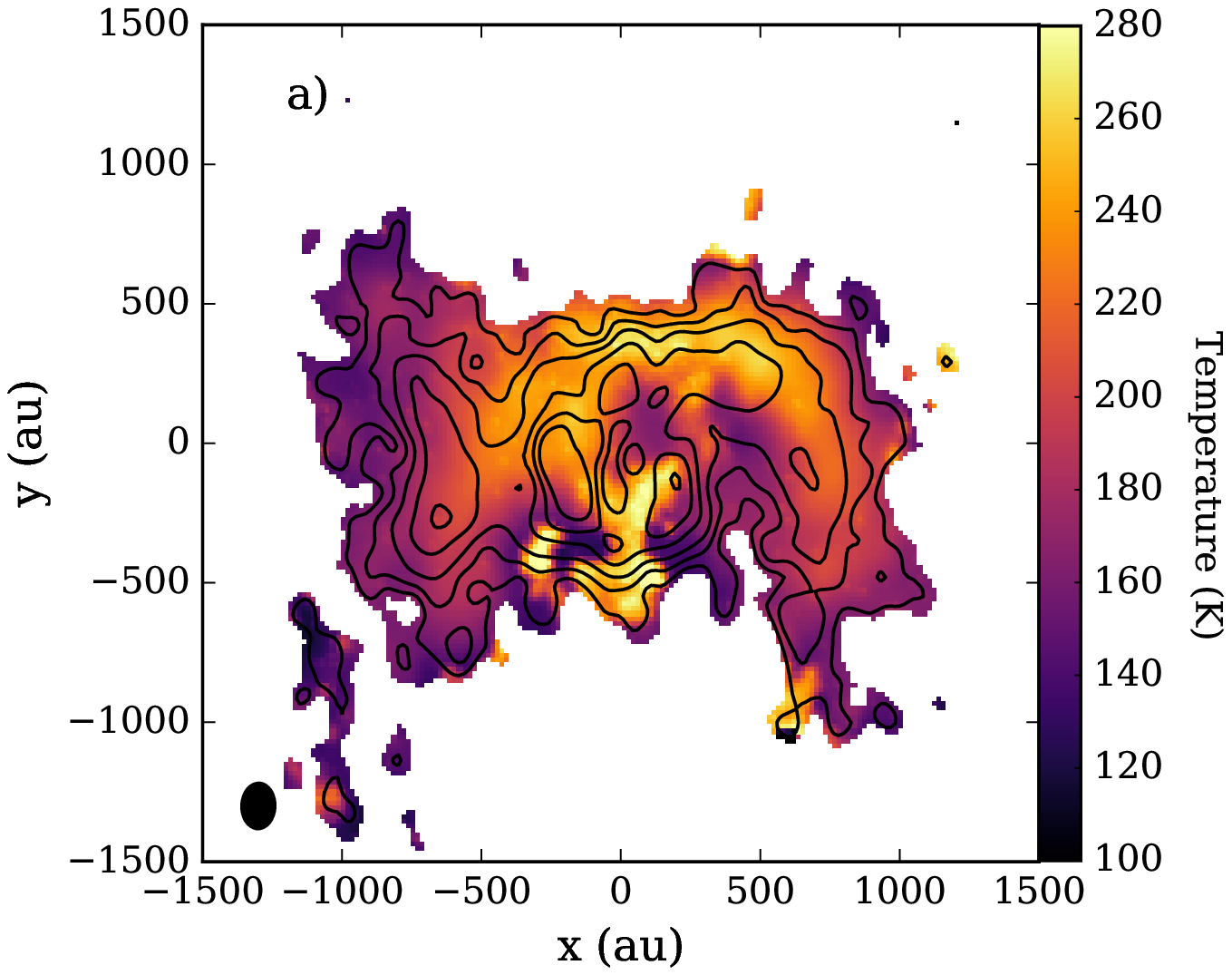}
            \includegraphics[width=8.3cm]{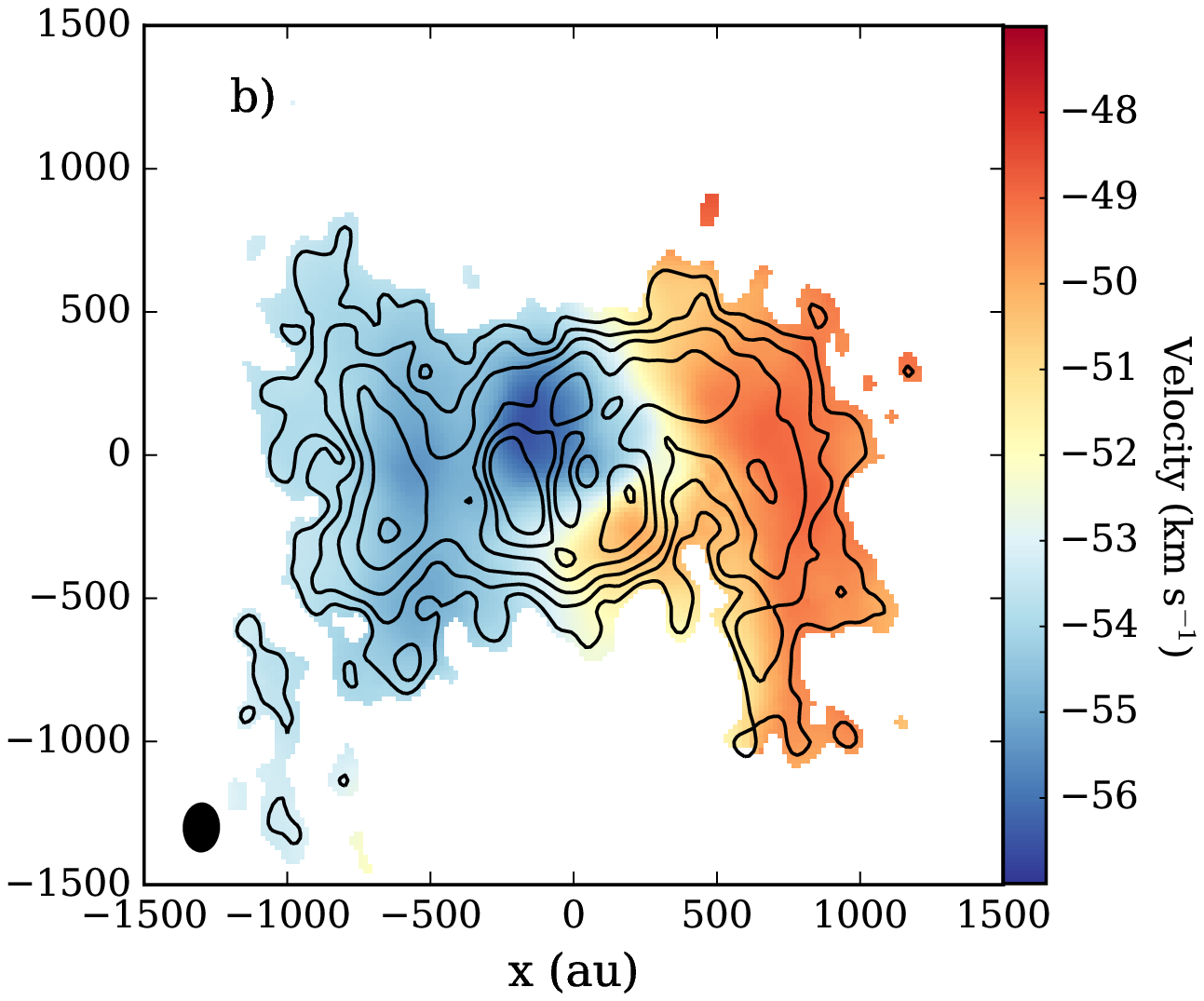}}
      \caption{Results of CASSIS modelling. Panel a: excitation temperature. Panel b: observed velocity. Both maps have been deprojected similarly to Fig.~\ref{zoomcont} (we note that the velocities are not yet corrected for inclination). The 1.21\,mm continuum emission as shown in Fig.~\ref{zoomcont} is overplotted as contours in both panels. The maps are masked to values above 4$\sigma$ in the continuum map. The deprojected beam is shown in the bottom left corner of each panel.}
         \label{cassis_temp_vel_fig}
   \end{figure*}

Figure~\ref{ch3cn_fig} presents the first moment map of the CH$_3$CN J=13-12 K=3 line emission from AFGL\,4176, which is overlaid with contours of 1.21\,mm continuum emission. The velocity gradient across the disc, which was originally seen by J15, is apparent in the data. However, further details can now also be seen, such as a velocity gradient across the inner part of the redshifted spiral arm. On the larger scales, the CH$_3$CN emission shows a bridge of gas between mm1-main and mm1c, which extends in a north-west direction from the blueshifted side of the disc. Therefore, the gas emission from the AFGL\,4176 region shows evidence of tidal interactions between mm1-main and possible companions, such as mm1c. The direction of this extended north-west arm or bridge mirrors the redshifted spiral on the west side of the disc, indicating mm1c may be exciting the spiral structure within mm1-main. An alternative explanation is that the process of ejection of mm1c from the disc could produce a stretched gas link between the secondary fragment or protostar and the primary.

Similar to the analysis carried out by J15, we created excitation temperature, velocity, linewidth, and column density maps across AFGL\,4176 by fitting the CH$_3$CN and CH$_3^{13}$CN J=13-12 K-ladder emission associated with each pixel with the Markov Chain Monte Carlo $\chi^2$ minimisation algorithm within the Centre d'Analyse Scientifique de Spectres Instrumentaux et Synth\'etiques tool (CASSIS)\footnote{CASSIS is currently being developed by IRAP-UPS/CNRS (http://cassis.irap.omp.eu).}. We used the same setup as described in J15 but with slightly updated initial parameter ranges ($T_{\rm ex}$ = 50 -- 350\,K, v$_{_{\rm LSR}}$ = -58 -- -48\,km\,s$^{-1}$, v$_{_{\rm FWHM}}$ = 0.5 -- 15\,km\,s$^{-1}$, and $N_{\rm mol}$ = 1$\times$10$^{15}$ -- 1$\times$10$^{18}$\,cm$^{-2}$). We determined the excitation temperature $T_{\rm ex}$ and velocity across the disc, as shown in Fig.~\ref{cassis_temp_vel_fig}. As we assume local thermodynamic equilibrium (LTE) in our analysis, $T_{\rm ex}$ therefore provides an estimate of the kinetic temperature. Maps of the remaining parameters (linewidth and column density) are presented as Figs.~\ref{cassis_ncol} and \ref{cassis_linewidth}. The maps in Fig.~\ref{cassis_temp_vel_fig} are deprojected and are clipped to the 4$\sigma$ level in the continuum maps (shown as contours).

Figure~\ref{cassis_temp_vel_fig}a shows that the temperature is lower towards the arc of continuum emission on the left, but higher in the inner leading section of the western spiral arm on the right, and directly to the upper left of the central source. We note that increased temperatures within spiral arms are seen in many simulations of self-gravitating discs \citep[e.g.][]{boley08a}. Figure~\ref{cassis_temp_vel_fig}b shows the velocity gradient across the entire disc as well as a velocity gradient across the inner part of the disc within several hundred astronomical units. 

\subsection{Resolved Toomre Q map}
Similar to the analyses conducted by \citet{ahmadi18a} and \citet{maud19a}, we used the results from previous sections to construct a map of the Toomre parameter $Q$ across the disc, given in Eq.~(\ref{toomre_eqn}). The sound speed $c_s$ can be calculated from the temperature, using the equation
\begin{equation}
\label{cs}
c_s = \sqrt{\frac{k T}{2.37 m_p}},
\end{equation}

\noindent where $k$ is the Boltzmann constant, $T$ is the temperature found from CASSIS, $m_p$ is the mass of a proton, and 2.37 is the mean molecular weight per free particle \citep{kauffmann08a}. The angular velocity $\Omega$ was found by subtracting the systematic velocity ($v_{\rm lsr}$=-52.5\,km\,s$^{-1}$) from the velocity map determined by CASSIS, deprojecting the velocities, and then dividing by the radius $r$ at each point in the disc to convert from tangential to angular velocities. Here we assume that the disc is close to Keplerian so that $\kappa=\Omega$; however, we have calculated $\kappa$ and the resulting map of $Q$ without this assumption in Appendix \ref{appendix_kappa}. Finally, the surface density $\Sigma$ can be calculated using the equation
\begin{equation}
\label{sigma}
\Sigma = \frac{g S_{\rm peak}}{\Theta B({\nu,T}) \kappa_{\nu}}
\end{equation}

\noindent where $g$ is the total-mass-to-dust ratio \citep[154$\pm15\%$, derived from the ratio of the total mass to mass of hydrogen of 1.402 and the ratio of the mass of dust to mass of hydrogen of 0.0091 from Tables 1.4 and 23.1 of][]{draine11}, $S_{\rm peak}$ is the peak flux of the ALMA continuum with the central compact source removed, $\Theta$ is the ALMA continuum beam area, $B(\nu,T)$ is the black body function, and $\kappa_{\nu}$ is the opacity \citep[0.24\,cm$^2$\,g$^{-1}$ at 1.21\,mm for $R_V=5.5$,][]{draine03a,draine03b}. Little is known about the properties of dust around massive forming stars, but as they form quickly, we assume that the dust is typical of star forming clouds, represented by the Draine Milky-Way $R_{\rm V}$=5.5 model.

\begin{figure}
\centering
\includegraphics[width=8.7cm]{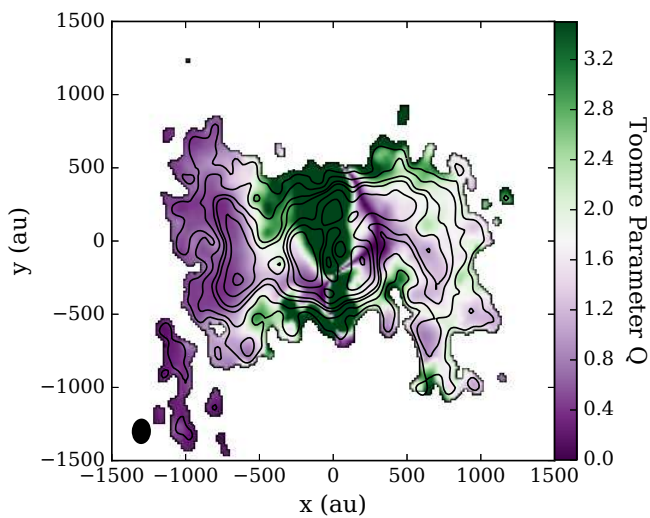}
  \caption{Deprojected map of $Q$, similar to that shown for Fig.~\ref{cassis_temp_vel_fig}. The low $Q$ values in the centre of the disc are an artefact caused by the velocity quickly shifting from blue to redshifted.}
     \label{toomre_map_fig}
\end{figure}

Figure~\ref{toomre_map_fig} shows the resulting map of $Q$ across the disc. The purple areas in the map show parts of the disc which are Toomre unstable \citep[Q$<$1.5-1.7,][]{durisen07a}. The low $Q$ values in the centre of the disc are due to an artefact caused by the velocity quickly shifting from blue to redshifted. However, further out in the disc we can see that the left (eastern) arc is clearly Toomre unstable, which may already contain a fragment at its lower end, as well as the densest parts of the western spiral on the right. Thus, our analysis shows that this disc has the right conditions to undergo fragmentation that may produce further companions.

We note that the CH$_3$CN does not trace the midplane, but hotter material higher in the disc atmosphere. In addition, although the disc is not globally optically thick, it may be locally optically thick, which would mean the surface density in the densest parts of the disc is underestimated. These optically thick parts of the disc would have to be unresolved, as the $<$1\,mJy\,beam$^{-1}$ emission from the outer disc has a brightness temperature of $<$20\,K. Hence by comparison with Fig.~\ref{cassis_temp_vel_fig}a, the outer disc is not optically thick at our resolution. From Eqs.~(\ref{toomre_eqn})-(\ref{sigma}) we can see that lowering the temperature and raising the surface density act to lower $Q$. Therefore, in this case, our analysis provides an upper limit on $Q$ and thus a lower limit on the level of instability of the disc.

In the case the assumed dust opacity $\kappa_{\nu}$ is a factor of two higher, this would mean that only parts of the disc in Fig.~\ref{toomre_map_fig} with $Q<0.75-0.85$ would be unstable. However, these values still occur in the left (eastern) side of the disc.

\section{Conclusions \label{conclusions}}
We present high-resolution ($\sim$30\,mas) 1.2\,mm ALMA observations of the AFGL\,4176 disc. We find that the disc is $\sim$1000\,au in radius and contains a spiral arm, which we fitted with logarithmic and Archimedean spiral models. The spiral structure is asymmetric and the observed CH$_3$CN gas emission shows evidence of gas arcs, which link AFGL\,4176 mm1-main to other millimetre sources in the field, providing evidence of interactions with companions. We produce a map of the Toomre parameter across the disc and find that the disc and spiral arms have already or are likely to fragment in future, possibly forming companions, thus showing that both current companions and gravitational instabilities play a combined role in creating structure in the AFGL\,4176 mm1-main disc. These results provide evidence that instability and fragmentation within discs is a possible pathway to explain the large number of multiple systems observed around high-mass stars.

\begin{acknowledgements}
We thank the referee for their comments which improved this manuscript, and John Ilee for useful discussions. AA and HB acknowledge support from the ERC under Horizon 2020 via ERC Consolidator Grant CSF-648505. H.B. also acknowledges support from the DFG in the Collaborative Research Center (SFB 881) ``The Milky Way System" (subproject B1). RK acknowledges support via his Emmy Noether Research Group funded by the DFG under grant no. KU 2849/3-1 and KU 2849/3-2. PB acknowledges support by the Russian Science Foundation (grant 18-72-10132).
\end{acknowledgements}

\bibliographystyle{aa}
\bibliography{} 

\begin{appendix} 

\section{Measuring the geometry of the mm1-main disc from continuum emission}

\begin{figure}
\centering
\includegraphics[width=9cm]{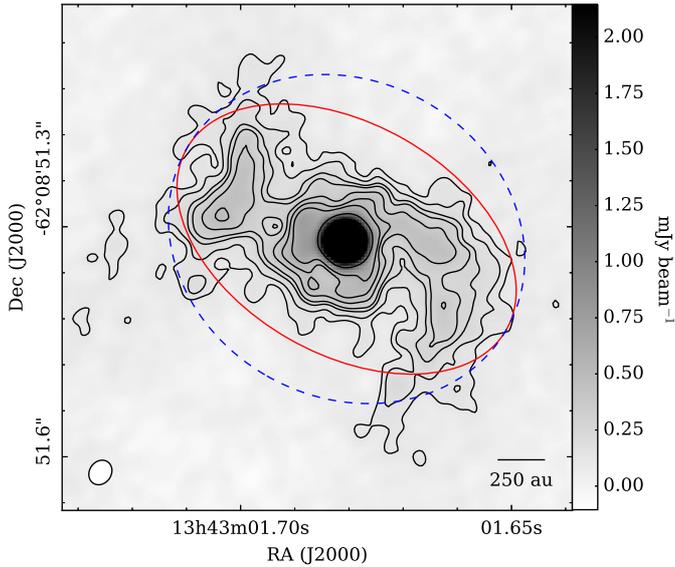}
  \caption{1.21\,mm continuum emission of AFGL\,4176 mm1-main in greyscale and contours ($\sigma$ = 23\,$\mu$Jy\,beam$^{-1}$ $
 \times$ -5, 5, 7, 10, 12, 16, 20, 25, 50, 100). The beam is shown in the bottom left corner (0.034$'' \times$0.028$''$, PA=-33.5$^{\circ}$), and a scalebar is in the right corner. Two ellipses with semi-major axes of 1000\,au or $\sim$0.24$''$ are shown, one with a semi-minor axis of 0.15$''$ (solid red, corresponding to $i=51^{\circ}$) and the other with a semi-minor axis of 0.22$''$ (dashed blue, corresponding to $i=30^{\circ}$).}
     \label{ellipse}
\end{figure}

 Figure~\ref{ellipse} shows a zoom-in of the 1.21\,mm continuum emission from mm1-main with ellipses used to measure its inclination over plotted (cf Section~\ref{contin_sec}). The red ellipse shows the smallest reasonable semi-minor axis of 0.15$''$, which corresponds to $i=51^{\circ}$, whereas the blue dashed ellipse shows the geometry expected for $i=30^{\circ}$.

\section{Maps of column density and linewidth from CASSIS modelling \label{extra_cassis}}

\begin{figure}
\centering
\includegraphics[width=9.3cm]{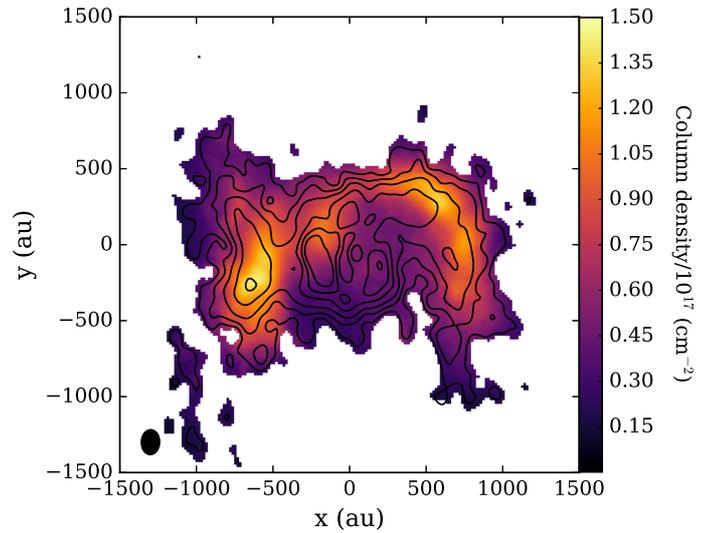}
  \caption{Results of CASSIS LTE modelling similar to Fig.~\ref{cassis_temp_vel_fig} but for column density.}
     \label{cassis_ncol}
\end{figure}

\begin{figure}
\centering
\includegraphics[width=9cm]{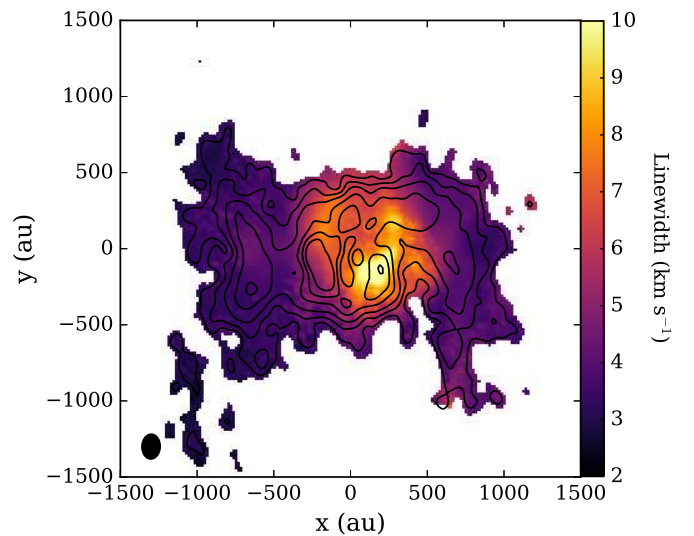}
  \caption{Results of CASSIS LTE modelling similar to Fig.~\ref{cassis_temp_vel_fig} but for linewidth.}
     \label{cassis_linewidth}
\end{figure}

Figures~\ref{cassis_ncol} (column density) and \ref{cassis_linewidth} (linewidth) show the remaining maps resulting from the CASSIS LTE modelling. The column density shown in Fig.~\ref{cassis_ncol} is highest in the leading section of the right or western spiral arm, which is offset from the spiral peak seen in the continuum, as well as in the lower end of the left or eastern arc, towards the continuum peak within the arc. The linewidth shown in Fig.~\ref{cassis_linewidth} shows a ring of high values around the position of the subtracted continuum source, which peaks to the lower right.

\section{Calculation of the Toomre parameter without assuming Keplerian rotation \label{appendix_kappa}}
Here we determine $\kappa$, the epicyclic frequency without the assumption of Keplerian rotation and use this to determine a new map of $Q$. The equation relating $\kappa$ to $\Omega$ in this case is 
\begin{equation}
\label{kappa}
\kappa^2 = \frac{2 \Omega}{r} \frac{d}{dr}(r^2 \Omega),
\end{equation}

\noindent where $r$ is the radius within the disc. Inserting $\Omega = v/r$ we get
\begin{equation}
\label{kappa2}
\kappa^2 = \frac{2 v}{r^2} \frac{d}{dr}(rv).
\end{equation}

\begin{figure}
\centering
\includegraphics[width=8.7cm]{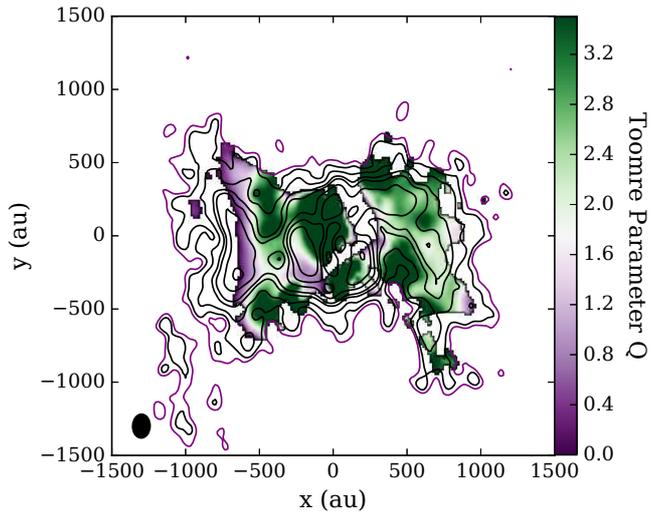}
  \caption{Deprojected map of $Q$ without the assumption that $\kappa=\Omega$. Undefined areas of the figure are enclosed by a purple contour; these regions are gravitationally unstable.}
     \label{toomre_kappa}
\end{figure}

To determine $\kappa$ across the disc, we used the velocity map shown in Fig.~\ref{cassis_temp_vel_fig} after subtracting the systematic velocity and deprojection of the velocities. The differential in Eq.~(\ref{kappa2}) was determined by finding the radial gradient of $rv$ across the map. The resulting value for $\kappa$ was inserted into Eq.~(\ref{toomre_eqn}) to determine the Toomre parameter $Q$. A map of $Q$ for the above assumptions is shown in Fig.~\ref{toomre_kappa}. 

In some regions of the disc, the derivative in Eq.~(\ref{kappa2}) is negative and therefore $\kappa$ is undefined, leading to gaps in the map. The physical meaning of a negative $\kappa^2$ is that radially moving a parcel of gas does not lead to oscillation, but instability. As $\kappa^2$ tends to zero, velocity shear has a reduced balancing effect against gravity, and as $\kappa^2$ goes negative, the velocity shear contributes to rather than suppresses instability. Thus, regions where $Q$ cannot be computed are unstable. To indicate this, we have drawn a purple contour around the area in which $\kappa^2$ was calculated. Our map of $Q$ (Fig.~\ref{toomre_kappa}), which was computed using $\kappa$, indeed shows in the eastern region that values of $Q$ start to show purple areas indicating instability, before being undefined, indicating strong instability. Our results are therefore consistent with the map derived from the assumption that $\kappa=\Omega$, which indicates that the eastern (left) region was unstable, although in this $Q$ map the western (right) spiral is stable.

\end{appendix}

\end{document}